\documentclass[prl,twocolumn,showpacs,amsmath,amssymb,showkeys]{revtex4-1}[11]

\usepackage{graphicx}
\usepackage{dcolumn}
\usepackage{bm}
\usepackage{wrapfig}
\usepackage{graphicx}
\usepackage{xcolor}
\usepackage[latin1]{inputenc}
\usepackage{graphicx}
\usepackage{graphics}
\usepackage{amsmath}
\usepackage{amssymb}
\usepackage{amsfonts}
\usepackage{mathrsfs}
\usepackage{latexsym}

\newcommand\R{\mathbb{R}}

\renewcommand\d\delta
\newcommand\D\Delta

\newcommand\beq{\begin{equation}}
\newcommand\eeq{\end{equation}}

\usepackage{amsfonts}
\usepackage{relsize}
\usepackage{graphicx}
\usepackage{dcolumn}
\usepackage{bm}
\usepackage{wrapfig}
\usepackage{graphicx}
\usepackage{url,lineno}
\usepackage{amsfonts}
\usepackage{color}
\usepackage[english]{babel}

\makeatletter

\begin{document}

\author{G. Florio}
\affiliation{Dipartimento di Meccanica, Matematica e Management, Politecnico di Bari, Via E. Orabona
4, I-70125 Bari, Italy}
\affiliation{INFN, Sezione di Bari, I-70126 Bari, Italy}

\author{G. Puglisi}
\affiliation{Dipartimento di Scienze dell'Ingegneria Civile e dell'Architettura, Via Re David 200, 700126,
Politecnico di Bari, Italy}

\date{\today}


 \title{Unveiling the influence of device stiffness
 in single macromolecule unfolding}


\begin{abstract}
\noindent Single-molecule stretching experiments on  DNA, RNA, and other biological macromolecules opened up the possibility of an impressive progress 
in many fields of Life and Medical sciences. The reliability of such experiments may be crucially limited by the possibility of determining the influence of the apparatus on the experimental outputs. Here we deduce an analytical model that we show to be coherent with  previous numerical results and that quantitively reproduce AFM experimental tests on titin macromolecules and P-selectin AFM experiments with variable probe stiffnesses. We believe that the obtained analytical description can represent an important step forward in the interpretation of Single Molecule Force Spectroscopy experiments and intermolecular interactions phenomena.

\end{abstract}

\maketitle



\section{Introduction}\label{sec:intro}

The comprehension of the role of mechanical forces at the molecular level represents nowadays the focus of incredible efforts in many different research fields of Biology, Biomechanics, Material Engineering, Biomedical Sciences. Fundamental phenomena such as DNA and RNA hairpins unfolding 
and refolding in enzymatic activity, sensing of metabolites, transcription termination and attenuation, morphogenic phenomena, cell motility and focal adhesion and so on cannot be described without a clear model of the underlying mechanical fields and macromolecular and cells force-response \cite{ING}.   
  
In recent years the possibility for the experimental study of protein and cells mechanical responses by 
Single Molecule Force Spectroscopy (SMFS) gave a great impulse in the understanding of the effects of mechanical forcees in the natural environment based on high-precision instruments such as Atomic Force Microscopes (AFM), optical tweezers, magnetic tweezers, microneedles. 
Differently from classical temperature or chemical denaturant based manipulation techniques, SMFS allows for the choice of specific trajectories in complex {\it bumpy} energy manifolds, with the possibility of analyzing the relative stability of locally stable configurations and topological properties of the entire bumpy energy landscapes \cite{EB, HS, WAS}. 
An important example is observed in the folding and refolding phenomena of the RNA or DNA secondary structure underlying gene transcription. Indeed in this case SMFS experiments show that unzipping propagates through a single stranded/unstranded front (diblock behavior) \cite{VT} that is not observed for temperature of chemically induced denaturation. Moreover SMFS experiments are clearly fundamental for the many proteins that are designated to withstand or transmit forces such as cytoskeletal or muscle protein macromolecules. 

One of the main drawbacks of SMFS is that the observed force-end to end distance diagrams are strongly influenced by unavoidable effects due to the pulling device. Although a significant effort both in a theoretical and in an experimental  perspective have been pursued on \cite{MA, MWL}, these effects are often neglected in many theoretical approaches and underestimated in the experimental field. In particular the two main aspects to be analyzed in this perspective are rate of loading effects and probe stiffness effects.

The theory proposed in this paper is based on equilibrium Statistical Mechanics so that our results can be applied to the rate-independent regime. This corresponds to the assumption that the relaxation rate to the energy minimum of the molecular chain is much higher than the external loading rate. 
In this perspective we point out that
two different behaviors have been typically observed \cite{LJ, FN}. At high rate of loading  a rate dependent  regime is observed and Kramer type relations are adopted with unfolding forces (logarithmly) growing with the loading rate \cite{DHS}. At low rate of loading ($k v
<  10$ pN, where $k$ is the instrument stiffness and $v$ the pulling rate \cite{LJ})
constant force thresholds are observed 
(quasistatic regime) with the force influenced from the only stiffness $k$. 
This rate-independent regime characterizes many SFMS experiments as well as many unfolding \cite{KB} and detaching ligand-receptor phenomena in physiological regimes \cite{LJ}.

Under our assumption of rate independence the main experimental artifact in SMFS experiments can be ascribed to the device stiffness. 
As reported {\it e.g.} in \cite{BMW} different experimental devices can show force transducer stiffnesses differing of five order of magnitudes  (from $10^{-3}$ to 100 N m$^{-1}$). 
Two ideal extreme cases can be considered depending on the ratio between the cantilever and the molecule stiffness \cite{KPL, MA}.  The first theoretical analysis case based on a Statistical Mechanics approach considering the probe energy was delivered in \cite{KPL} where the authors model the molecule as a chain of elements with convex potential energy in series with an harmonic springs with variable stiffness describing the probe effect. Numerical examples to describe the experimental behavior of poly (ethylene glycol) molecules have also been analyzed.

In the case of stiff cantilevers (hard device) the system can be described  using the Helmholtz ensemble: the molecule is held at a fixed extension with the corresponding force representing a `fluctuating' dual variable. In the opposite case of soft cantilever  (soft device), the Gibbs ensemble is adopted with an assigned force acting on the molecule and a fluctuating end-to-end length.  The two ensemble can be shown to be equivalent in the thermodynamical limit with the Helmholtz and Gibbs free energies correlated by a Legendre transform  \cite{MGP}. 

Even more important is the effect of probe stiffness when one considers, as we do in this paper, molecules constituted by elements exhibiting two (or more) `stable' configurations. As we show in this paper, completely different experimental responses can be observed for the same macromolecule under such different experimental boundary conditions. For example in the fundamental case of protein macromolecules constituted by two state (folded/unfolded) modules, by varying the probe stiffness the behavior can range from force-plateaux to sawtooth force-elongation diagrams corresponding to a cooperative and a non cooperative transition, respectively \cite{MGP,PTa,PTb}. This results also in important variations of the force inducing the dissociation of a molecular complex \cite{FMG} the unfolding of a protein \cite{Rief} or of higher-order protein structure \cite{CB}.

From a theoretical point of view this corresponds to the non simple extension of the theory in \cite{KPL} to chains of elements with non convex potential energies.  In the following, to fix the ideas, we focus (in the experimental validation and in the nomenclature) on systems of elements undergoing conformational transitions between a folded (e.g. $\alpha$-helices and $\beta$-sheets structures) and unfolded 
state as widely observed in stretching experiments of biomolecules. On the other hand, systems with non convex (free) energies have been successfully adopted in several fields: the application of Landau theory for the description of the appearance of spontaneous magnetization in the ordered phase (below the critical temperature) \cite{Ker} or the martensitic phase transformation of shape-memory alloys \cite{martensitic} are well known successful examples of this approach. The obtained results can then be important also in several other different fields.

It is important to remark that also the case of chains with non convex energies has been extensively analyzed. 
In particular we recall that the possibility of deducing the hard and soft cases as limit models for a system with variable probe stiffness, was numerically shown in \cite{KPL} in the deduction of a model predicting the behavior of single polyethylene glycol chains AFM experiment.The analysis is more subtle when one considers local energy minimizers (metastable configurations)  that can survive also in the thermodynamical limit \cite{PT5}. 
The case of a chain with two-wells energy, in the limit of hard device, has been studied in \cite{Lev2010}. In particular, in order  to obtain an analytical expression of the partition function and analyze the thermodynamical limit, the authors use the approximation that we adopt here to extend the quadratic energy wells beyond the spinodal point. The transition to the soft regime has been studied using Monte Carlo numerical simulations in \cite{Manca}.

On the other hand in real experiment one cannot expect a regime completely described by the hard or soft device limits. On the contrary, one would like to have a theory describing and predicting the behavior of the unfolding phenomenon in an intermediate regime. In this paper we precisely address this issue and move forward in this field. We are able to obtain explicit expressions for the partition function and force-strain relations that are valid in the \emph{whole} range of values of the measuring device stiffness. 

The comparison with AFM pulling behavior on titin macromolecule and P-selectin single module unfolding at different probe stiffnesses demonstrate the effectiveness of the model in describing the experimental response of 
real biological molecules with particular accurate predictivity.

\section{Results}\label{sec:results}

\begin{figure}[t]
\begin{center}
\includegraphics[width=0.45\textwidth]{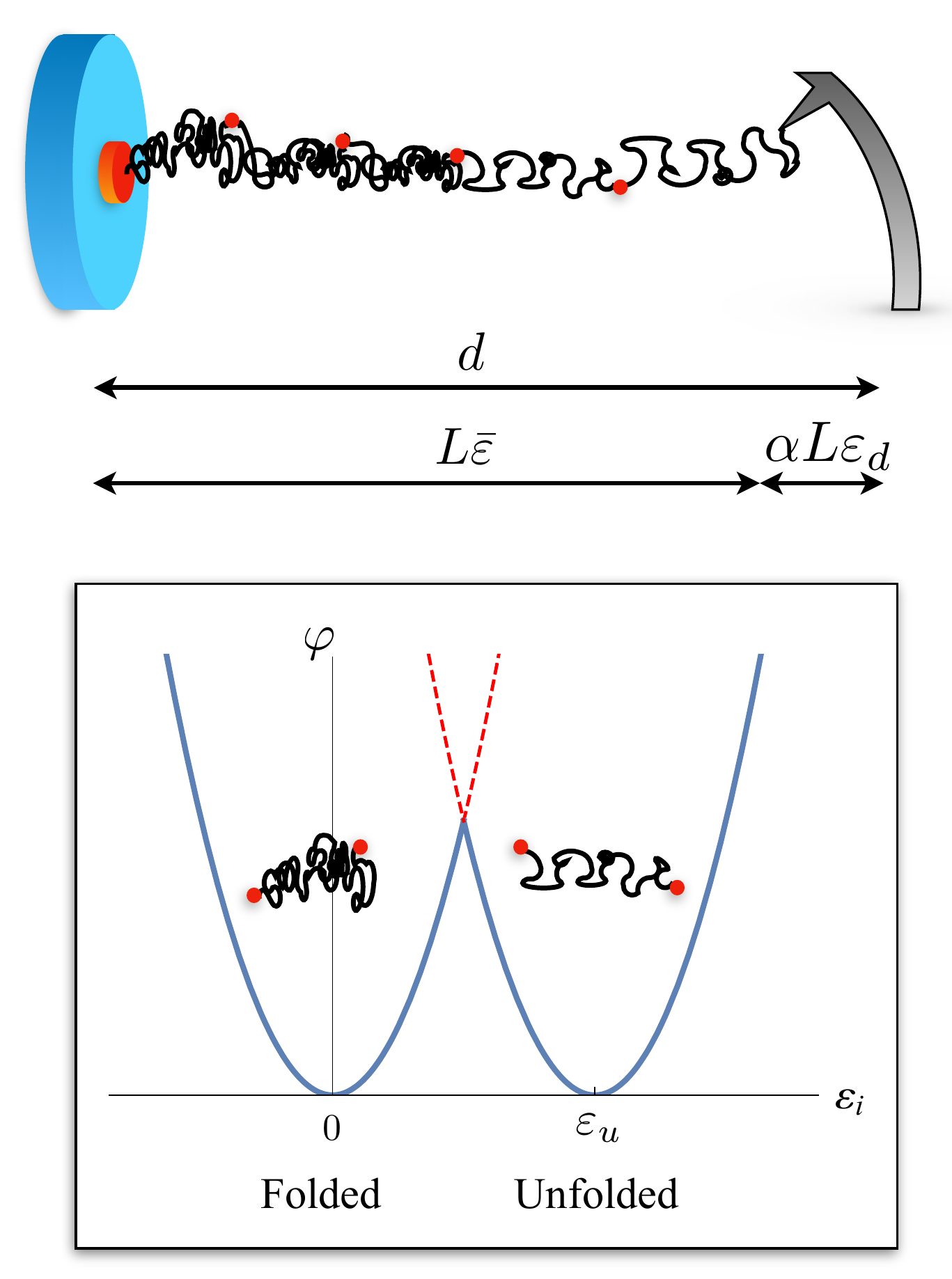}
\caption{Scheme of an SFMS experiment for the analysis of the unfolding of a protein molecule constituted by modules undergoing a folded/unfolded transition here modeled through a two well potential energy assumption.}
\label{fig:scheme}
\end{center}
\end{figure}

Consider an AFM experiments on a molecule constituted by bi-stable elements. To fix the idea we refer to the important case of a protein molecule (such as titin \cite{Rief}) constituted by ($\alpha$-helix or $\beta$-sheets) folded elements undergoing a conformational transition (unfolding) due to the elongation imposed by a pulling device
 (see the scheme in Fig.\ref{fig:scheme}).
We model the macromolecule as a chain of $N$ masses connected by bistable springs and a probe connected to the $N$-th oscillator through a spring with stiffness $k_d$. We want to analyze an experiment with assigned displacement $d$.

The potential energy of the chain is
$
V_c=\sum_{i=1}^{N}\frac{L}{N} \varphi(\varepsilon_i),
$
where $L$ is the end to end length of the unloaded molecule. Since the folded/unfolded transition of protein folded domains is typically an all or none transition we assume for each domain a two well energy
$
\varphi(\varepsilon_i)= \frac{k_p}{2} (\varepsilon_i -\varepsilon_u\chi_i)^2.
$
Here $\varepsilon_i$ is the strain of the $i$-th spring, $\chi_i$ is a `spin variable'
such that $\chi_i=0$ if the $i$-th spring is in the first (folded) state and $\chi_i=1>0$ if it is in the second (unfolded) state and $\varepsilon_u$ is the reference strain of the second well. Observe that to get analytical results we approximate the energy wells by parabolic laws and that for simplicity of notation we assume that the spring stiffness $k_p$ is the same in the two states. We assume also that the transition (Maxwell) force of the springs is zero. The following analysis is easily extended to the case when these hypotheses are not assumed. In particular fully analytical results can be attained when the two well energy is approximated by a two parabolic wells with different stiffnesses and non zero Maxwell force \cite{PTa}.

The total (molecule plus loading device) potential energy is obtained adding the energy $V_d$ of the measurement device. For small displacements ($\sim 150$ nm) we can approximate the probe as a linear spring \cite{NN}.
We thus have
\begin{equation}
V=V_c+V_d=\sum_{i=1}^N \frac{L}{N} k_p \frac{(\varepsilon_i -\varepsilon_u\chi_i)^2}{2}
+ \alpha L k_d \frac{\varepsilon_d^2}{2}\label{V}
\end{equation}
where $\varepsilon_d$ is the probe deformation and $\alpha L\varepsilon_d$ is its length. 
The assigned displacement is then related to the strains by the following relation:
\begin{equation}\label{eq:constraint}d=\sum_{i=1}^N \frac{L}{N}\varepsilon_i+ \alpha L \varepsilon_d
=L (\bar \varepsilon+ \alpha \varepsilon_d)\end{equation}
where we have introduced the average chain strain
\begin{equation}\label{eq:avestrain}
\bar \varepsilon=\frac{1}{N}\sum_{i=1}^N \varepsilon_i.
\end{equation}

\paragraph{Zero temperature.---}\label{sec:zerotemp}

Consider first the zero temperature limit
when the equilibrium configurations are characterized by a constant force $F$ with
\begin{equation}\varepsilon_i=\frac {F}{k_p}+\varepsilon_u\chi_i \,\,\,\, i=1,...,N, \,\,\,\, \varepsilon_d=
\frac {F}{k_d}.\label{eq}\end{equation}
We then obtain the force-strain relation
\begin{equation} 
F=k_p ( \bar \varepsilon -\varepsilon_u \, {\bar \chi}),\label{aeq}
\end{equation}
where 
$
{\bar \chi}:=\displaystyle \frac{1}{N}\sum_{i=1}^N \chi_i \in[0,1]
$ is the fraction of unfolded elements.

By using \eqref{eq:constraint}-\eqref{aeq}  we then obtain that at equilibrium
\begin{equation}
\bar \varepsilon=
\gamma \delta+(1-\gamma)\varepsilon_u \, {\bar \chi},
\end{equation}
where $\delta=d/L$ and 
\begin{equation}\label{eq:gamma}
\gamma=\frac{k_d}{k_d+ \alpha k_p}\in \,\,  ]0,1[ \,\, ,
\end{equation}
is the {\it main non dimensional parameter of the paper} measuring the relative stiffness of the probe {\it vs} the full  (chain plus probe) system. The ideal limits of `hard' and`soft' devices correspond to $\gamma \rightarrow 1$ and  $\gamma \rightarrow 0$, respectively.

\begin{figure*}[t]
\begin{center}
\includegraphics[width=0.7\textwidth]{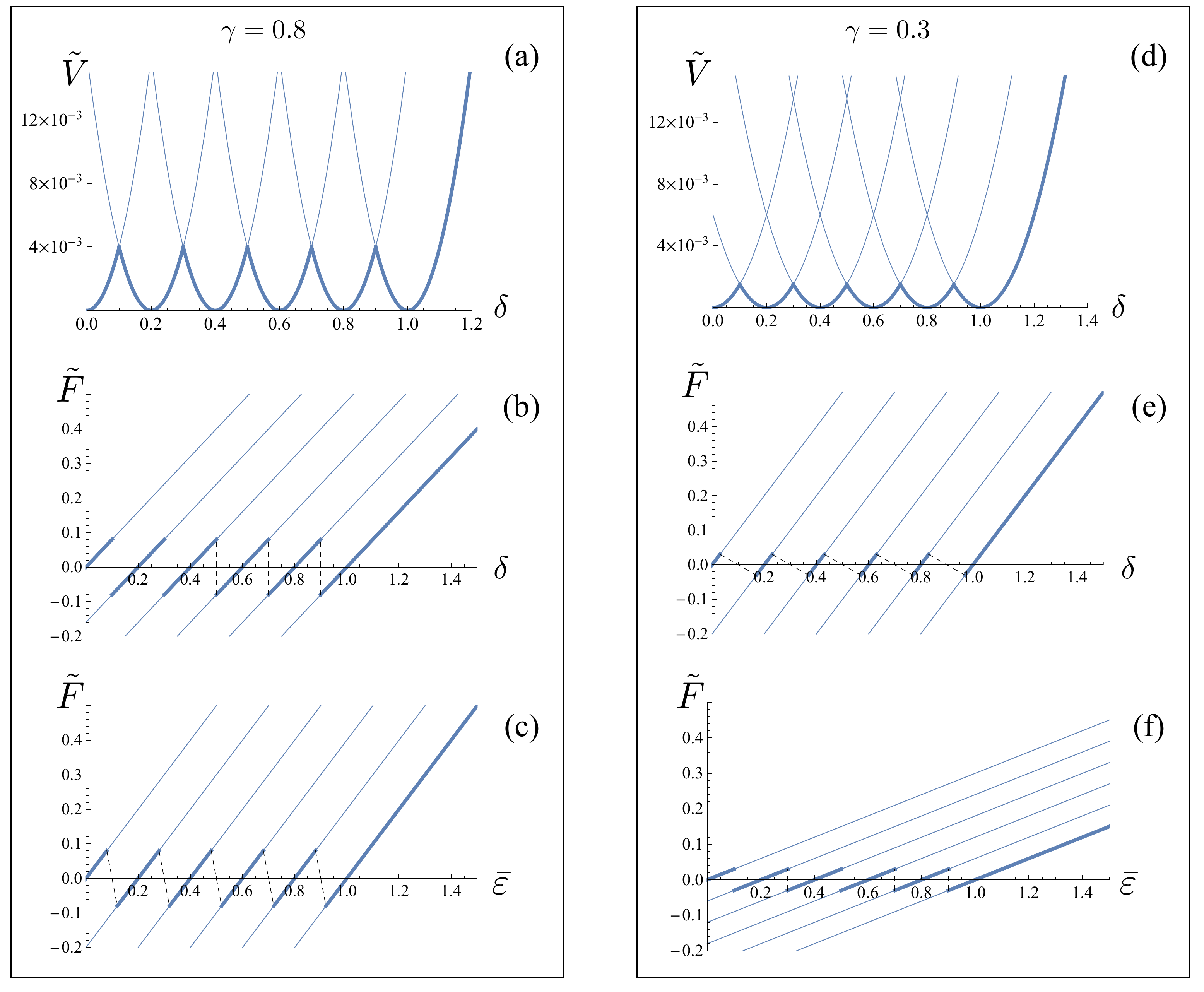}
\caption{Plot of $\tilde{V}$, $\tilde{F}(\delta)$ and $\tilde{F}(\epsilon)$ at zero temperature for a chain with $N=5$ elements and $\varepsilon_u=1$: $\gamma=0.8$ [(a), (b), (c)] and $\gamma=0.3$ [(d), (e), (f)]
}
\label{fig:zerotemp}
\end{center}
\end{figure*}
Using \eqref{V} and \eqref{aeq} it is possible to obtain the (non dimensionalized) force $\tilde F=\frac{F}{k_p}$
and potential energy $\tilde V=\frac{V}{k_p L}$ for the generic equilibrium configuration at given unfolded fraction $\bar \chi$:
\begin{equation}\begin{array}{l}
\tilde F(\delta)=\gamma (\delta-\varepsilon_u\, {\bar \chi}),\vspace{0.2 cm} \\
\tilde V(\delta)=\frac{\gamma}{2} (\delta-\varepsilon_u \, {\bar \chi} )^2.\label{Veq}
\end{array}\end{equation}
Observe that all two-phase solutions with ${\bar \chi} \in ]0,1[$ are defined only for $\| \tilde F \| \leq \varepsilon_u$. Thus, the existence domain of each equilibrium branch is \begin{equation}
\delta \in \left[\varepsilon_u \, {\bar \chi}-\frac{1}{\gamma}, \varepsilon_u \, {\bar \chi}+\frac{1}{\gamma}\right ].
\end{equation}
The fully folded state is defined for $\delta\leq \frac{1}{\gamma}$ whereas the fully unfolded configuration for 
 $\delta \geq \varepsilon_u- \frac{1}{\gamma}$.

One can deduce that all these solutions are metastable (local energy minima) due to the local convexity of the energy of both springs and probe. The global minima of the energy can be evaluated using \eqref{Veq}. The two-phase configurations with phase fraction ${\bar \chi}$ corresponds to the global minimum for 
\begin{equation}
\delta \in \left [\varepsilon_u \, {\bar \chi}-\frac{1}{2N}, \varepsilon_u \, {\bar \chi}+\frac{1}{2N } \right].
\end{equation}
The fully folded state represents the global minimum for 
$\delta\leq \frac{1}{2N}$
whereas the fully unfolded configuration corresponds to the global minimum of the energy for  $\delta \geq \varepsilon_u-\frac{1}{2N}$.
We thus obtain the fundamental result at zero temperature: the unfolding force is constant and it depends 
on the relative stiffness of the probe with respect to the macromolecular chain. In particular, if we increase $\delta$ from the fully folded state then the chain starts to unfolds at the threshold
\begin{equation}\label{eqFu}
F_{u}=\frac{\gamma}{2N}.
\end{equation}
On the other hand, if we let the system refold by relaxing the probe and let $\delta$ decreases,
the chain start to refold to go back to the primary folded state at
the threshold
\begin{equation}
F_{f}=-\frac{\gamma}{2N}.
\end{equation}
Observe that these values should be considered as incremental force from the transition force of the bistable spring, here assumed null for simplicity of analytical expressions.

In figure \ref{fig:zerotemp} we represent the important modifications of the obtained unfolding behavior depending on the relative stiffness parameter $\gamma$.  \vspace{0.3 cm}

\paragraph{Non-zero temperature.---}\label{sec:finitetemp}

In this section we analyze the fundamental temperature effect on the unfolding molecular behavior at different values of $\gamma$ in the framework of equilibrium Statistical Mechanics \cite{Ker}. 

Following the suggestion in \cite{Lev2010}, to get analytical result, we expand the two energy wells for the potential energy on the real line, beyond the spinodal point where they intersect. In figure \ref{fig:scheme} we have marked these regions with the dashed branches of the parabola. It is possible to show (see the Supplementary Information) that in the typical temperature range of 
living proteins this approximation is fully acceptable because the `artificial' configurations with higher potential energy do not modify significantly the partition function.

Standard calculations (see SI) deliver
the canonical partition function $Z_N$ of a chain of $N$ elements \emph{and} the measuring apparatus using the approximations of our model: 
\begin{equation}\label{eq:partitionfunctext}
Z_N=K_N{\left(1-\gamma\right)}^{1/2}\sum_{p=0}^N \left( \begin{array}{c} N \\ p \end{array}\right)e^{-\frac{\beta k_p l\gamma N}{2}{(\varepsilon_u\frac{p}{N}-\delta)}^2},
\end{equation}
where $K_N$ is a constant, $\beta=1/k_B T$, $k_B$ is the Boltzmann constant, $T$ is the absolute temperature and  the binomial coefficient gives the number of configurations of the chain with fraction $p/N$ of unfolded domains
($\bar \chi=p/N$). 

Observe that the terms in the sum are gaussian functions centered in $\varepsilon_u \,p/N$ and width proportional to $1/\sqrt{\beta k_p l\gamma N}$. At low temperature (large $\beta$) the Gaussian terms are squeezed and contributions to $Z_N$ are relevant only for values of $\delta$ very close to $\varepsilon_u\, p/N=\varepsilon_u\, \bar \chi$ with $p=0,\dots,N$. As expected this limit reproduce the behavior described in previous section based on the minimization of the total potential energy [see 
 Eq.\eqref{Veq}].

The free energy is defined as
\begin{equation}\label{eq:freeen}
\Phi_N=-\frac{1}{\beta}\ln{Z_N}
\end{equation}
and the expectation value of the $F$ conjugate to the displacement $\delta$ is
\begin{equation}\label{eq:sigmadef}
 \langle F \,\rangle =\frac{1}{L}\frac{\partial}{\partial \delta}\Phi_N=-\frac{1}{\beta L}\frac{1}{Z_N}\frac{\partial}{\partial \delta}Z_N.
\end{equation}
From Eq.(\ref{eq:partitionfunctext}) we thus find
\begin{equation}\label{eq:sigma}
 \langle F \,\rangle =k_p\gamma \left(\delta-\varepsilon_u\, \langle {\bar \chi} \,\rangle  \right),
\end{equation}
where
\begin{equation}\label{eq:gammafunc}
 \langle {\bar \chi} \,\rangle =\frac{\sum_{p=0}^N \left( \begin{array}{c} N \\ p \end{array}\right)\frac{p}{N} e^{-\frac{\beta k_plN\gamma}{2}{(\varepsilon_u\frac{p}{N}-\delta)}^2}}{\sum_{p=0}^N \left( \begin{array}{c} N \\ p \end{array}\right)e^{-\frac{\beta k_p l N\gamma}{2}{(\varepsilon_u\frac{p}{N}-\delta)}^2}}
\end{equation}
Here $ \langle {\bar \chi} \,\rangle $ is the expectation value of the fraction $\xi=p/N$ of unfolded domains at fixed temperature $T$ and assigned displacement $\delta$. 
Similarly (see Supplementary Information) we obtain the expectation value of the average strain in Eq.(\ref{eq:avestrain})
\begin{eqnarray}\label{eq:averagestraintext}
 \langle \bar{\varepsilon} \,\rangle &=&\delta-\frac{1-\gamma}{k_p \gamma}\left(-\frac{1}{\beta}\frac{1}{L Z_N }\frac{\partial}{\partial \delta}Z_N\right)\nonumber\\ 
&=&\varepsilon_u\,\langle \bar \chi \,\rangle +\gamma\left(\delta-\varepsilon_u\,\langle \bar \chi \,\rangle \right).
\end{eqnarray}
Finally, from Eq.(\ref{eq:sigma}) and (\ref{eq:averagestraintext}) we obtain the \emph{central} result of the paper giving the force-strain relation at assigned temperature and relative stiffness $\gamma$
\begin{equation}\label{eq:finalsigma}
 \langle F \,\rangle =k_p\left( \langle \bar{\varepsilon} \,\rangle -\varepsilon_u\, \langle {\bar \chi} \,\rangle \right)
\end{equation}
as implicit functions of the assigned displacement $\delta$. Remarkably, we deduce that Eq.(\ref{eq:finalsigma}) represents the extension of the equilibrium equation Eq.(\ref{aeq}) to the case of non-zero temperature when the force-strain relation is regulated by the expectation value of the unfolded fraction $\bar \chi$. This equation fully characterizes the material response of the macromolecule for the whole range of stiffness ratio $\gamma$.

\begin{figure}[h!]
\begin{center}
\includegraphics[width=0.4\textwidth]{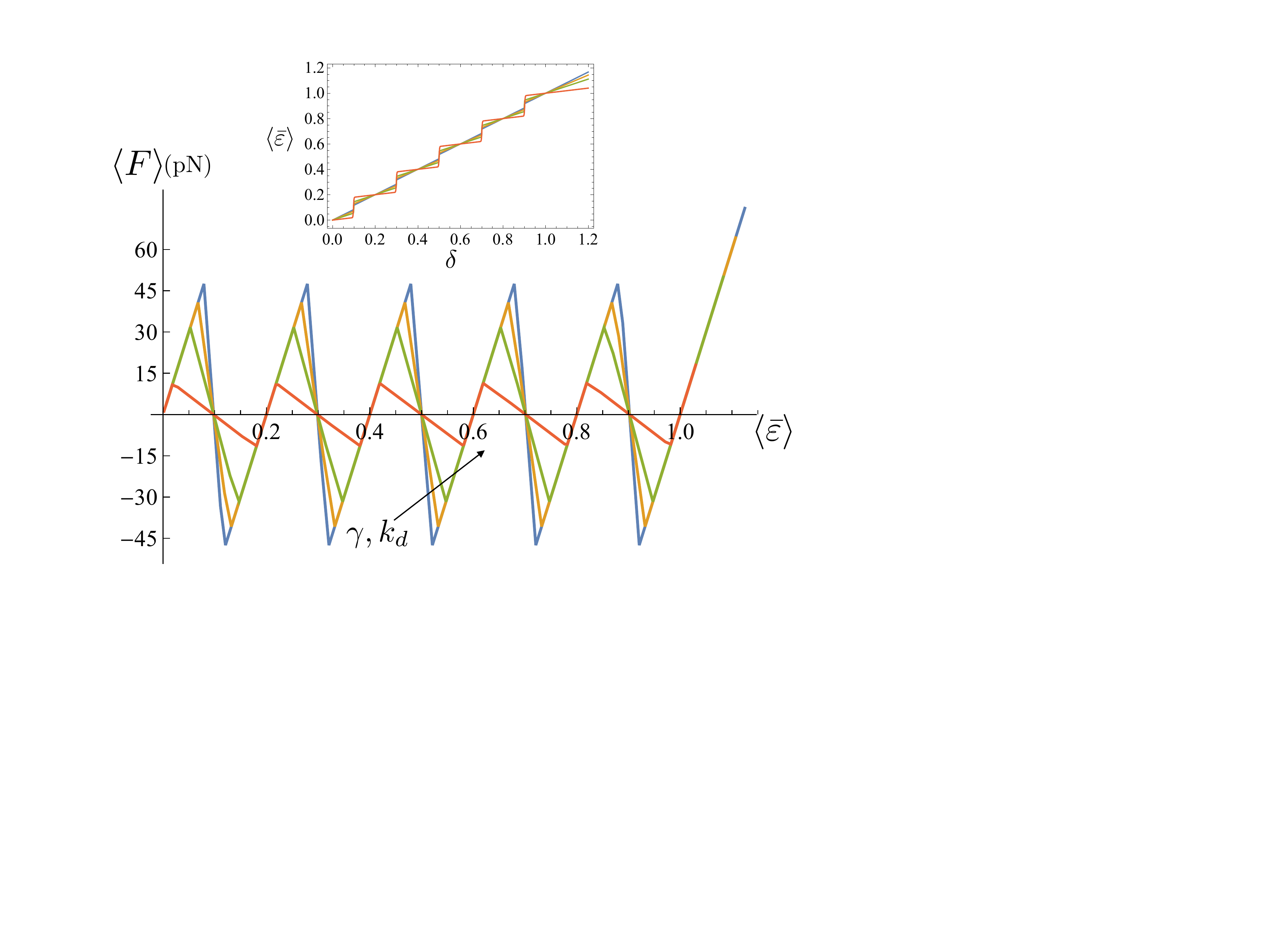}
\caption{Force-strain curves for different values of $\gamma$ (and $k_d$). The direction of the arrow denotes decreasing values of these quantities. We have fixed $N=5, \varepsilon_u=1, T=300\,\mbox{K}, l=30 \,\mbox{nm}$ and $\bar{k}_{exp}=k_p/Nl=4\,\mbox{pN/nm}$. The curves are obtained for $\bar{k}_d=k_d/(\alpha L)=20, 10, 5, 1\,\mbox{pN/nm}$ corresponding to  $\gamma\simeq 0.83, 0.71,0.56,0.2$, respectively (see the Supplemetary material). Inset: dependence of  $\langle \bar{\varepsilon}\rangle$  on $\delta$ (same values of the parameters in the larger figure).
}
\label{fig:force1}
\end{center}
\end{figure}

\begin{figure}[h!]
\begin{center}
\includegraphics[width=0.35\textwidth]{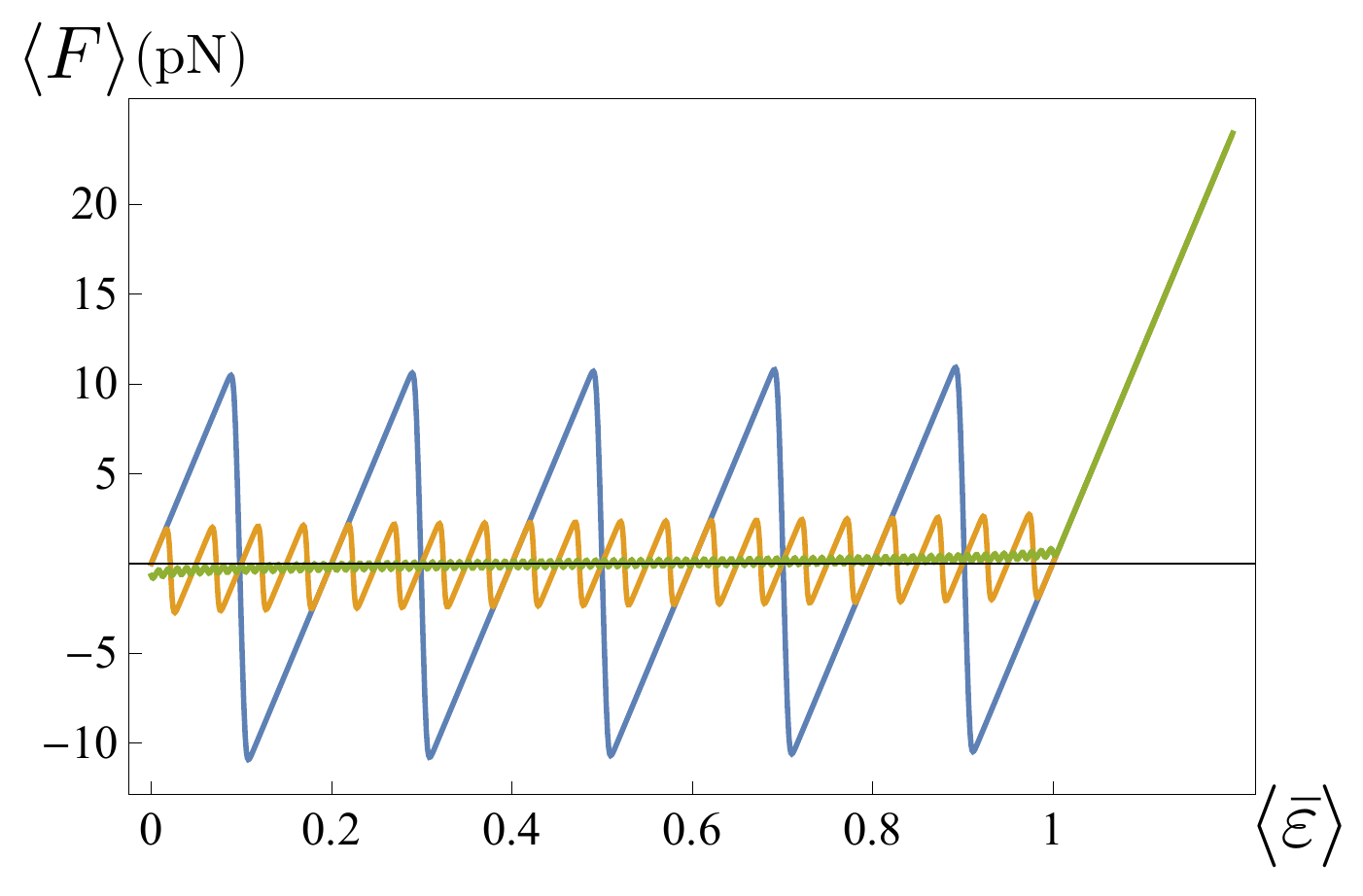}
\includegraphics[width=0.35\textwidth]{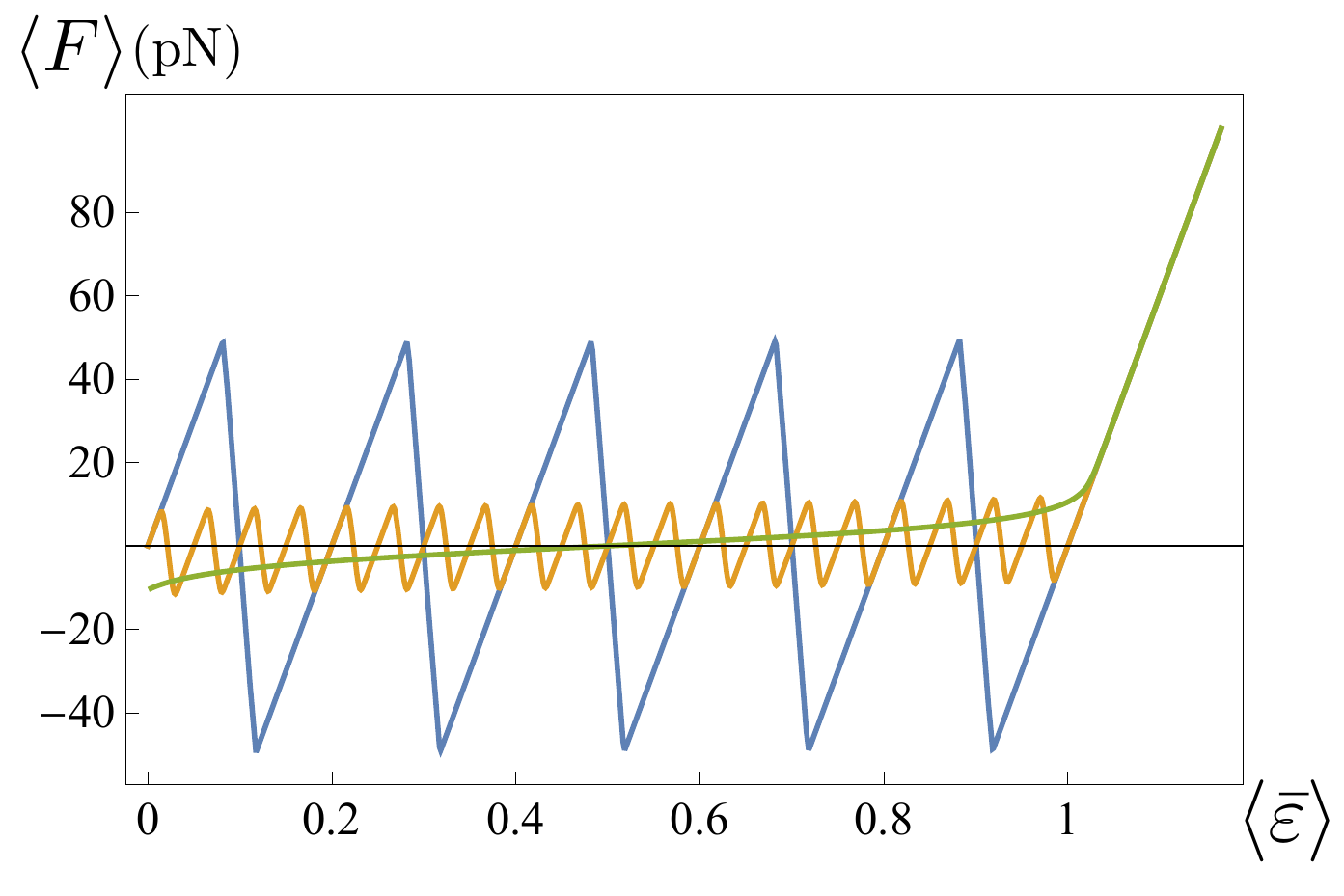}
\caption{Top: force-strain curve for $N=5, 20,100$ with $l=30 \,\mbox{nm}$ fixed. Larger values of $N$ correspond to smaller amplitude of the oscillations. We have fixed $\varepsilon_u=1, T=300\,\mbox{K}$ and $\bar{k}_p=k_p/l=4\,\mbox{pN/nm}, \bar{k}_d=k_d/(\alpha L)=20\,\mbox{pN/nm}$.\\
Bottom: force-strain curve for $N=5, 20,100$ with total length $L=Nl=150\,\mbox{nm}$ fixed. Larger values of $N$ correspond to smaller amplitude of the oscillations. We have fixed $\varepsilon_u=1, T=300\,\mbox{K}, \bar{k}_d=k_d/(\alpha L)=20\,\mbox{pN/nm}$ and $\bar{k}_{exp}=k_p/Nl=4\,\mbox{pN/nm}$ with $l=30\,\mbox{nm}$ ($N=5$), $l=7.5\,\mbox{nm}$ ($N=20$), $l=1.5\,\mbox{nm}$ ($N=100$).}
\label{fig:force2-3}
\end{center}
\end{figure}

\begin{figure}[h!]
\begin{center}
\includegraphics[width=0.35\textwidth]{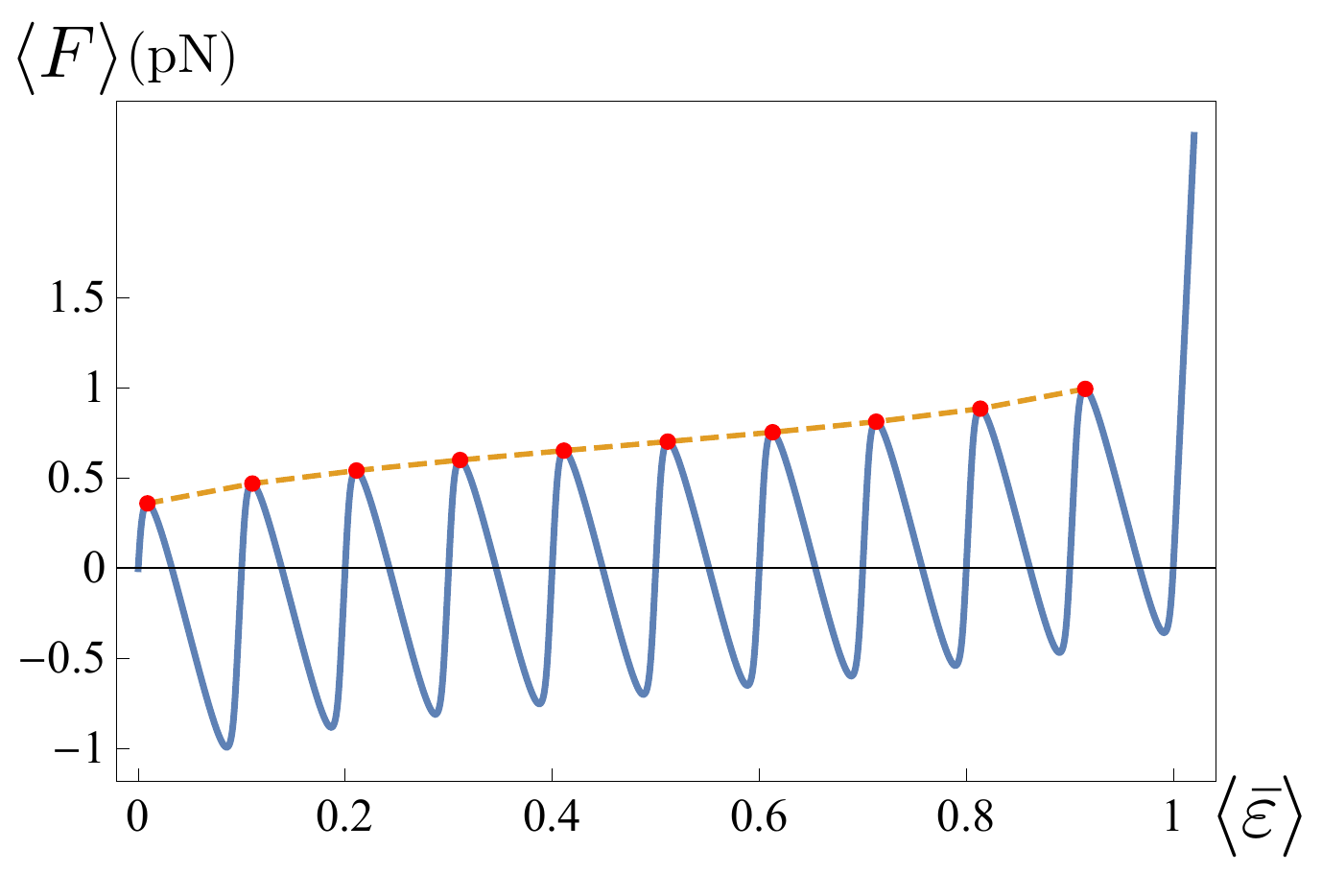}
\includegraphics[width=0.35\textwidth]{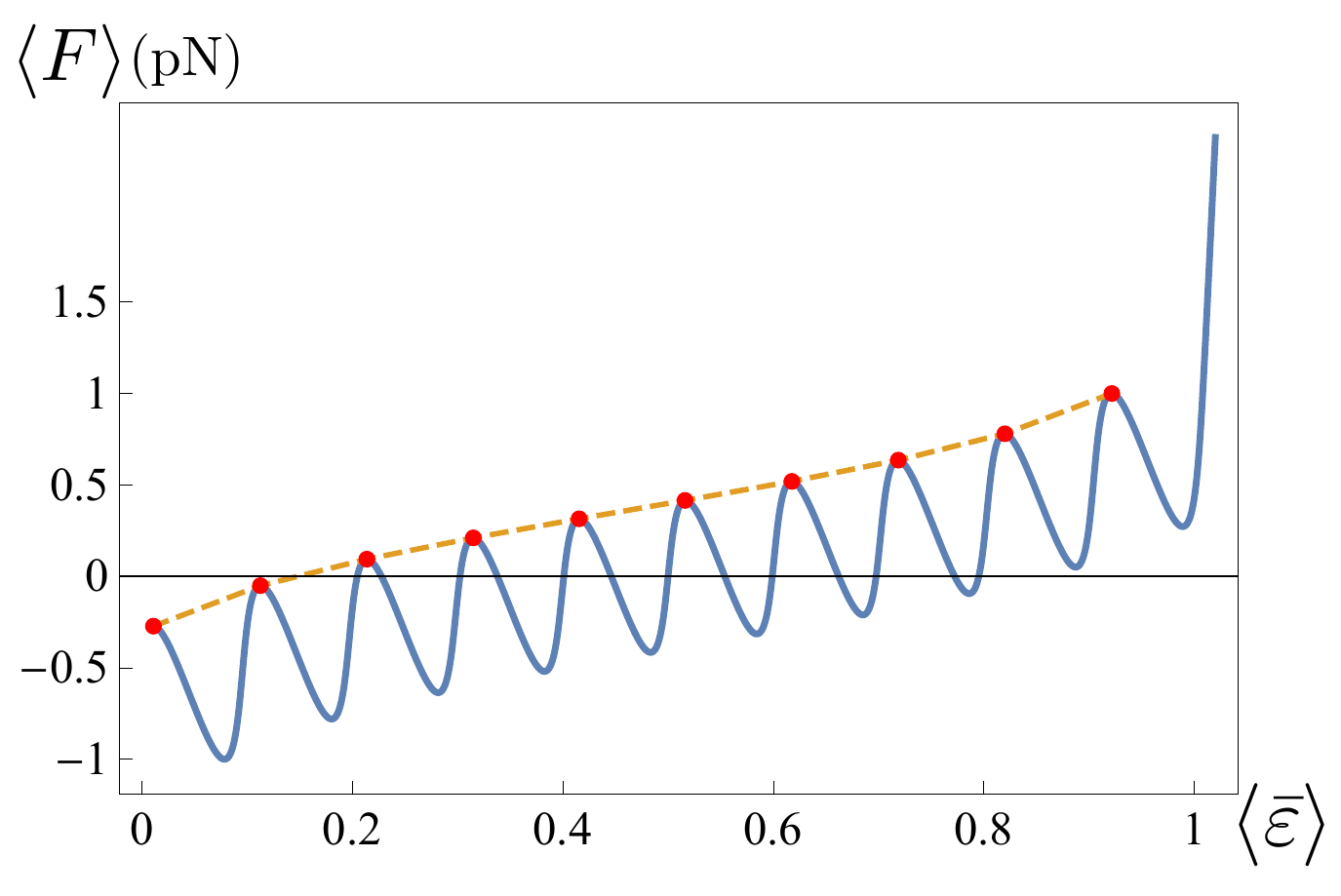}
\caption{Force-strain curve showing the hardening effect of  local maxima of $F$ increasing with $\langle \bar \varepsilon \rangle$. We have fixed  $N=10$, $\varepsilon_u=1, T=300\mbox{K}$ (top), $T=600\mbox{K}$ (bottom),  $l=30 \mbox{nm}$ and $\bar{k}_{exp}=k_p/Nl=0.4\mbox{pN/nm}, \bar{k}_d=k_d/(\alpha L)=0.1\mbox{pN/nm}$ (corresponding to $\gamma= 0.2$).}
\label{fig:force4}
\end{center}
\end{figure}

In Figure \ref{fig:force1} we plot the force-strain curve for fixed value on $N$, temperature $T$, $l=L/N$ and stiffness of an element of the chain $k_p$; we consider a set of curves obtained for different values of $\gamma$ corresponding to different stiffness of the probe $k_d$ (see the caption for details). It is important to observe the different shapes of the response curves determined depending from the parameter $\gamma$ and in particular that the unfolding threshold $F_u$, as in the zero temperature case \eqref{eqFu}, is a decreasing function of $\gamma$.

 In Figure \ref{fig:force1} we represent the evolution of the protein average strain $ \langle \bar \varepsilon \,\rangle $
as the applied displacement is increased.  Observe that for decreasing values of $\gamma$ (i.e. softer devices) each unfolding transition corresponds to growing strain discontinuities so that the strains grows with localized strain discontinuities. In the opposite case of hard devices the strain evolution is continuous.

In Figure \ref{fig:force2-3} we study the interesting effect of increasing $N$. As expected, the amplitude of the oscillations decreases with $N$. We notice that in one case (top figure) we have increased $N$ keeping $l$ fixed. On the other hand, in Figure \ref{fig:force2-3} (bottom) we have increased $N$ keeping $L=Nl$ fixed. According with this observation we deduce that the effect of stiffness variation is particularly important for small systems whereas it disappears in the thermodynamical system analyzed in the following section.

Finally, observe that according with the AFM unfolding experiments on bimolecular molecules such as titin \cite{Rief},  the values of the local maxima of $F$ are not constant, but increase with $\langle \bar{\varepsilon}\rangle$ (hardening, see Figure \ref{fig:force4}).  This effect can be interpreted observing that as $\langle \bar{\varepsilon}\rangle$ grows the number $p$ of unfolded domains is increased. Correspondingly  the number of  available
configurations with $p+1$ unfolded domains decreases. Interestingly, our  tests show that the hardening effect is more evident for higher temperatures (see Figure 5)  and larger number $N$ of domains (see Figure 4 bottom). This effect is important because in the case of assigned force corresponds to a transition from a cooperative to a non cooperative unfolding behavior.  In \cite{DMPS} this effect has been addressed to a possible inhomogeneity of the unfolding domain. It is important to remark that according with our results we can interpret it as a temperature effects. \vspace{0.3 cm}

\paragraph{Thermodynamical limit.---}\label{sec:tdlimit}

Many protein materials are constituted by macromolecules with a large number of $\beta$-sheets folded domains
undergoing conformational transformation under increasing end-to-end distance. It is interesting then
to consider the limit $N\rightarrow+\infty$. 

We begin by searching the asymptotic expression of the function $\langle{ \bar \chi}\rangle$ in Eq. (\ref{eq:gammafunc}).
Using the saddle point method (see SI for details) we find the expectation value of the unfolded fraction
\begin{equation}\label{eq:Gammaasymp}
\langle \bar \chi \,\rangle \sim
 \chi_c(\delta).
\end{equation}
where $\chi_c\in (0,1)$ is the minimum of the function 
\begin{equation}
f(x)=S(x)+\tilde{\beta}(\varepsilon_u x-\delta)^2 
\end{equation}
with 
\begin{equation}
S(x)=x\ln x+(1-x)\ln(1-x),
\end{equation}
and $\tilde{\beta}= \beta l k_p \gamma /2$.
Finally, we have
\begin{equation} \label{eq:tlcase1}
\langle F \rangle \sim
k_p\gamma \left(\delta-\varepsilon_u \chi_c\right)
\end{equation}
and
\begin{equation} \label{eq:tlcase2}
\langle \bar{\varepsilon}\rangle \sim
\varepsilon_u \chi_c+\gamma (\delta-\varepsilon_u \chi_c).
\end{equation}

\begin{figure}[h!]
\begin{center}
\includegraphics[width=0.35\textwidth]{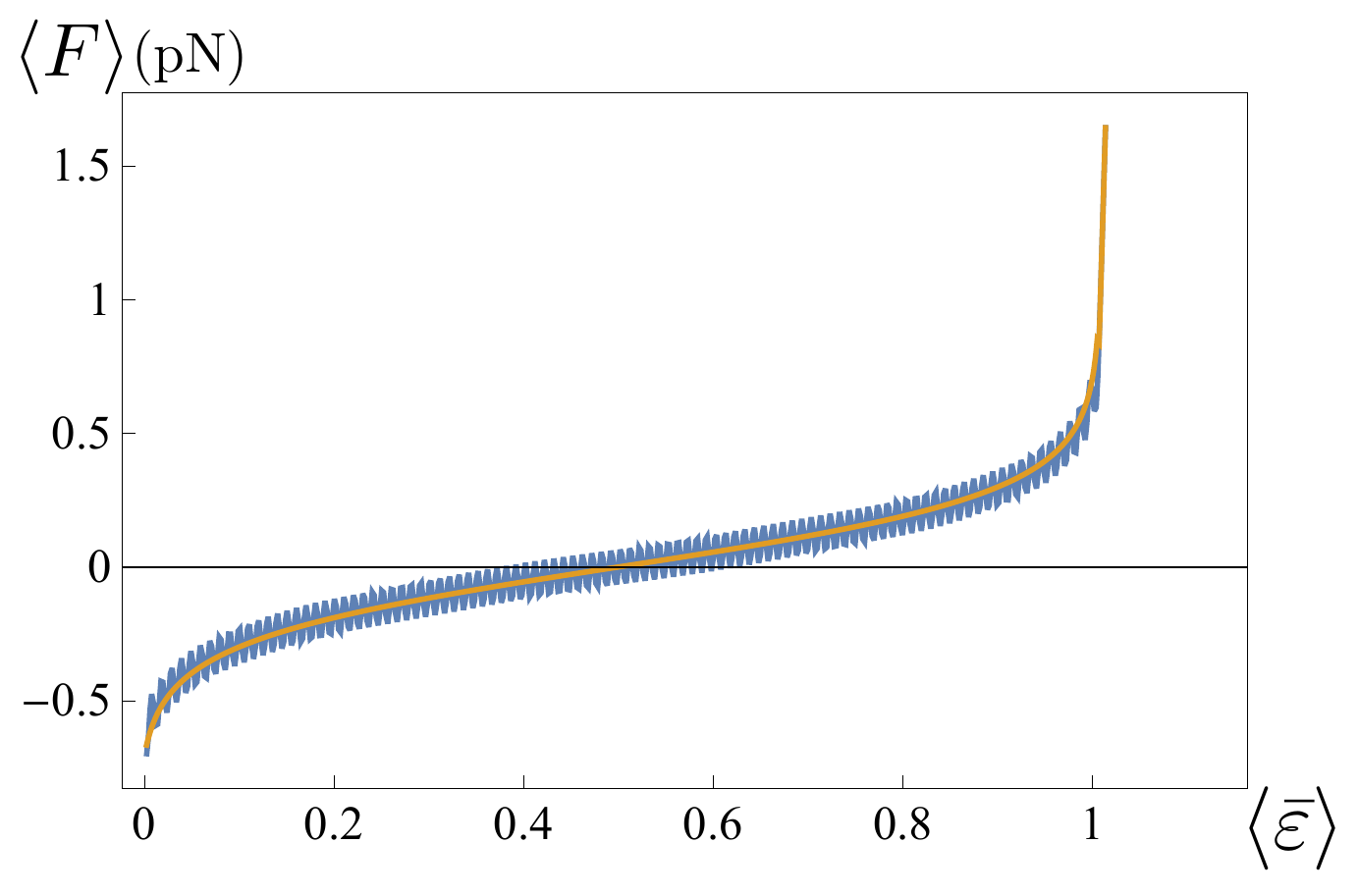}
\caption{Comparison of the force-strain curves obtained in the thermodynamical limit (monotonic curve) and for $N=100$ (chainsaw curve). We have fixed $\varepsilon_u=1, T=300\,\mbox{K}, l=30 \,\mbox{nm}$, $\bar{k}_p=k_p/l=4\,\mbox{pN/nm}, \gamma=0.7$.}
\label{fig:forcetl}
\end{center}
\end{figure}

In Figure \ref{fig:forcetl} we show the excellent agreement of the force-strain curves obtained for $N=100$ and the thermodynamical limit case Eq.(\ref{eq:tlcase1})-(\ref{eq:tlcase2}).\vspace{0.3 cm}

\paragraph{Comparison with experimental results.---}\label{sec:experiment}

\begin{figure}[t!]
\begin{center}
\includegraphics[width=0.28\textwidth]{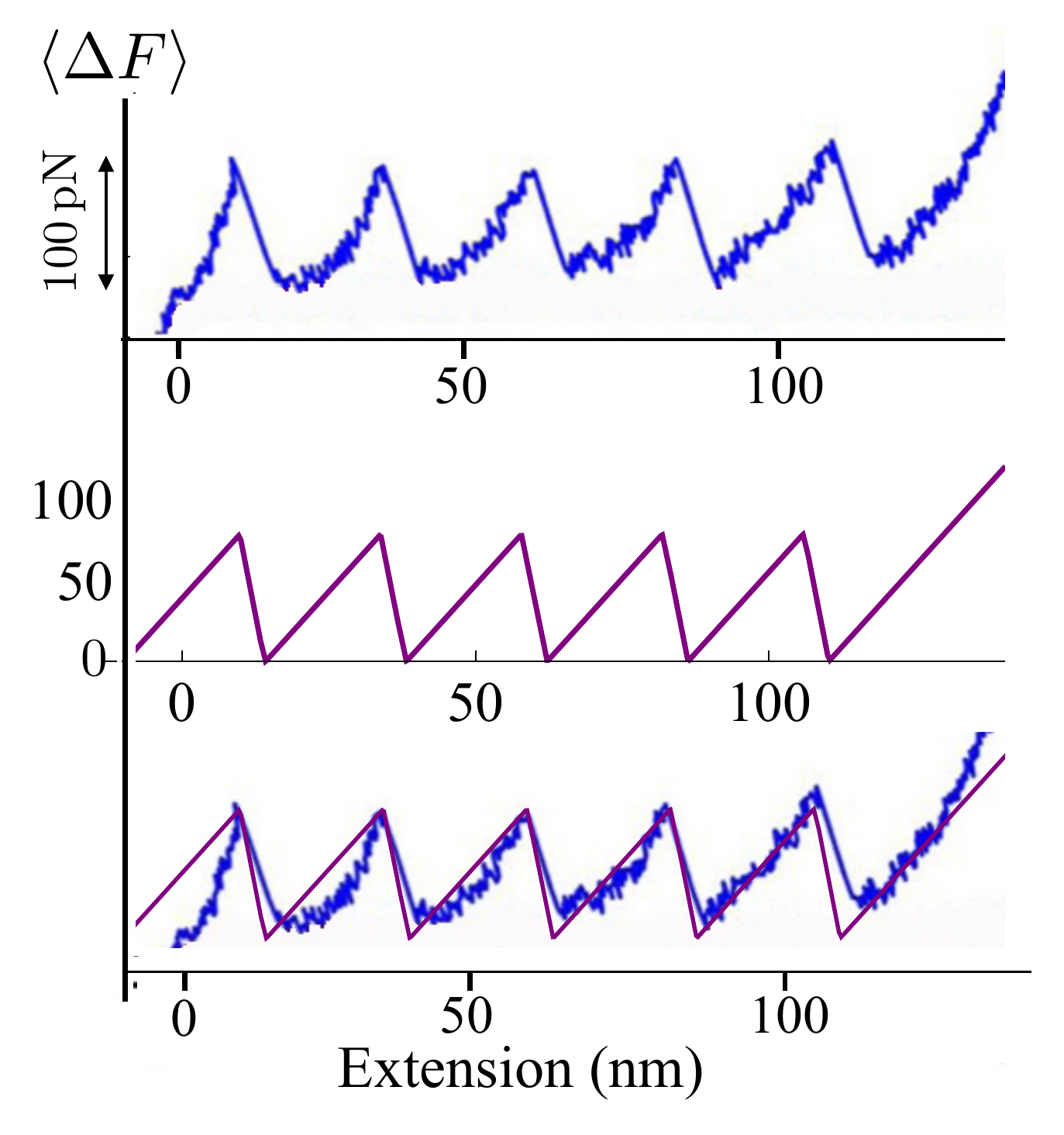}
\caption{Comparison of the the force-strain curves obtained from the model with the experimental results (see \cite{cardiac}). We have fixed  $N=5$, $\varepsilon_u=1, T=300\,\mbox{K}, l=24 \,\mbox{nm}$ and $\bar{k}_{exp}=k_p/Nl=4\,\mbox{pN/nm}, \bar{k}_d=k_d/(\alpha L)=20\,\mbox{pN/nm}$ (corresponding to $\gamma\simeq 0.86$).}
\label{fig:titin-confront1}
\end{center}
\end{figure}

In order to show the predictive capacities of the proposed model, in Figure \ref{fig:titin-confront1}  we compare the theoretical response with AFM experimental unfolding behavior of titin reported in \cite{cardiac}. In the theoretical curve we assign the spring constant of the AFM used in those experiment ($\simeq20\mbox{pN/nm}$) and we set the stiffness of the titin as $\bar{k}_{exp}=4\,\mbox{pN/nm}$ which is coherent with typical values of titin PEVK unfolding energies (see \cite{DMPS}, \cite{pevk} and SI). In the figure we have indicated $\langle \Delta F \rangle$ because in our model the Maxwell stress of the two-wells potential energy is set to zero. We remark that the average unfolding force can be easily calibrated on the experimental results by simply considering a non-zero Maxwell stress. 

\begin{figure}[h!]
\begin{center}
\includegraphics[width=0.28\textwidth]{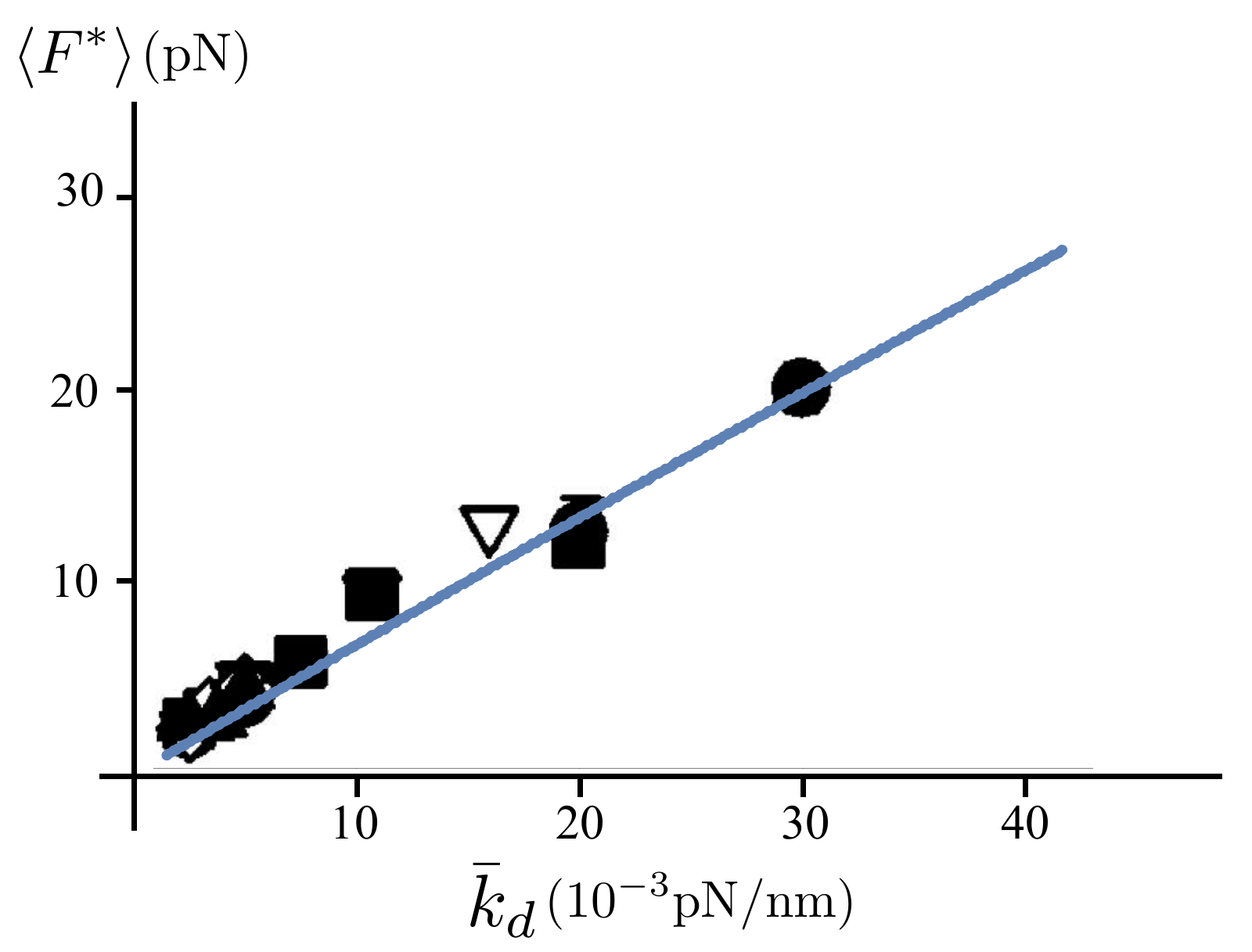}
\caption{Dependence of the rupture force on the spring constant of a (very soft) transducer compared with experimental results (see \cite{psgl1}). We have fixed  $N=1$, $\varepsilon_u=35, T=300\,\mbox{K}, l=40\, \mbox{nm}$ and $\bar{k}_p=1\,\mbox{pN/nm}, \bar{k}_d=k_d/(\alpha L)$ varying from $1\times 10^{-3}\,\mbox{pN/nm}$ to $40\times 10^{-3}\,\mbox{pN/nm}$. Different symbols correspond to different loading rates.}
\label{fig:explinear}
\end{center}
\end{figure}

To further test our results against the experimental behavior, in Figure \ref{fig:explinear} we
study the dependence of the rupture force  $\langle F^*\rangle$ for dissociating P-selectin-PSGL-1 bond as a function of the spring constant of a (very soft) transducer  as reported in \cite{psgl1}. To this scope, following \cite{fpt}, we describe the protein rupture as a transformation of a system with a two-wells potential energy and second minimum `far' from the first one, {\it i.e.} $\varepsilon_u\gg 1$. Thus, we identify the fracture force $\langle F^*\rangle$ with the maximum attained value of $\langle F\rangle$ as the rescaled total displacement $\delta$ is increased.  The values of the parameters are taken from \cite{psgl1} except $\varepsilon_u$ (fixing the energy barrier) whose value is chosen in order to fit the experimental results. The linear trend obtained from our model is consistent with the analysis performed in \cite{psgl1}.

\section{Discussion}\label{sec:discussion}

The comprehension of the response of macromolecules is fundamental in many fields of biology, medicine and bio-inspired materials engineering. Indeed, the incredible elasticity and recovery properties of bio-materials result from the features of the constituent macromolecules. In this paper, we have proposed a simple and effective description of the macromolecule behavior modeled as a chain  of elements undergoing conformational transitions (such as in $\beta$-sheets and $\alpha$-helices unfolding). SFMS techniques constituted in the last decades the main experimental tools to determine the undergoing energetics. As we show in this paper, the only way to interpret such experiments is to model the macromolecule and the measuring device as a unique system (both at zero and non-zero temperature). 
 Here, for the first time, we deliver explicit analytic solutions describing the force {\it vs} end-to-end distance diagram as a function of the main non-dimensional parameter measuring the relative device versus macromolecule's stiffness. As this parameter is varied the model describes the experimentally observed regimes going from sawtooth transitions in the case of isometric (hard device $\equiv$ $k_d\rightarrow \infty$) conditions to the cooperative type (force plateau) transition observed for the assigned force experiments (soft device $\equiv$ $k_d\rightarrow 0$). As well known \cite{Manca}, such an effect is often ignored in the interpretation of the experimental results. We believe that the approach here proposed and its possible extensions can represent an important step forward in the field. In particular, the presence of mechanical interactions for proteins, cells and biological tissues play a crucial role in so many diseases and biological functions \cite{goriely}, we observe that a direct extension of our approach to such problems can give a new  theoretical framework correctly considering the boundary conditions {\it e.g.} for biomaterial growth and medical pathologies. 

\vspace{0.1cm}

\paragraph{Acknowledgements.---}

We thank the authors of the papers \cite{cardiac} and \cite{psgl1} for the authorization to use their data in order to compare them with our model. We also thank M. Ligab\`{o} for useful comments. G. Puglisi is supported by the Prin Project 2015 Advanced mechanical modeling of new materials and structures, by Gruppo Nazionale per la Fisica Matematica  (GNFM) of the Istituto Nazionale di Alta Matematica (INdAM) and by FRA project of Politecnico di Bari. G. Florio is supported by GNFM (INdAM) through ``PROGETTO GIOVANI'', by INFN through the project ``QUANTUM'' and by MIUR through the FABBR research grant. 

\vspace{0.1cm}
\paragraph{Author contributions.---} G.F. and G.P. contributed equally to the paper.

\vspace{0.1cm}
\paragraph{Materials and Correspondence.---} 
Correspondence and requests for materials should be addressed to G.F. (giuseppe.florio@poliba.it) or G.P. (giuseppe.puglisi@poliba.it)

\vspace{0.1cm}
\paragraph{Additional information.---}
Supplementary Information accompanies this paper.

\clearpage

\section*{Supplementary Information for ``Unveiling the influence of device stiffness
 in single macromolecule unfolding''}\label{sec:suppmat}
 \renewcommand{\theequation}{SI\arabic{equation}}
\setcounter{page}{1}
\renewcommand{\thefigure}{SI\arabic{figure}}
\setcounter{figure}{0}

\setcounter{equation}{0}


\subsection{Non-zero temperature}

We introduce here the Hamiltonian function of the whole system constituted by the chain and the apparatus. We then obtain the resulting partition function, the free energy and the force-deformation equation. 

The Hamiltonian of the system can be written as
\begin{equation}\label{eq:hamiltonian}
H=E_k+V,
\end{equation}
where $E_k$ and $V$ take into account the contributions of kinetic and potential energies, respectively. In particular, we have 
\begin{equation}\label{eq:kinetic}
E_k=\sum_{j=1}^{N}\frac{N{p^2_i}}{2m}+\frac{{p^2_{N+1}}}{2M}
\end{equation}
where $p_j,\,j=1,\dots N+1$ are the linear momenta, $m/N$ and $M$ denote the mass of each element of the chain and the representative mass of the apparatus, respectively. 
On the other hand, $V$ contains both the non-convex potential energies of the bistable elements and the elastic contribution describing the interaction between the $N$-th element of the chain and the measuring apparatus. Thus, we have
\begin{equation}
V=\sum_{j=1}^{N} \Phi(\xi_{i})+\frac{\bar k_d}{2}{(\xi_{N+1})}^2
\end{equation}
where $\xi_i=x_i-x_{i-1}$ is the displacement difference between the elements $i$ and $i-1$ (with the convention $\xi_0$=0), $\xi_{N+1}=\xi_d$ as in Eq.(\ref{V})  and 
\begin{equation} \label{eq:potentialwell}
\Phi(\xi_{j})=\frac{\bar k_p}{2}{(\xi_j-a_j)}^2.
\end{equation}
The material constants $\bar k_p$ and $\bar k_d$ represent the stiffness of the bistable elements in the chain and of the measuring device, respectively. They are related to the constant in Eq. \eqref{V} of the main paper by the following relations:
\begin{equation} 
\bar k_p=k_p/(L/N)=k_p/l,\quad \bar k_d=k_d/(\alpha L).
\end{equation} 
Notice that $a_j$  can assume two different (non negative) values corresponding to the two different (folded and unfolded) minima $a_j^{(1)},a_j^{(2)}$. As a preliminary consideration, we notice that in the following we introduce a simplification, i.e. we expand the two energy wells in Eq.(\ref{eq:potentialwell}) on the whole real line, beyond the spinodal point where they intersect. In figure \ref{fig:scheme} of the main paper we have marked these regions with the dashed branches of the parabola. This simplifies the calculations and let us obtain analytical expressions.

The partition function in the canonical ensemble takes the form
\begin{widetext}
\begin{equation}\label{eq:partition}
Z_N=\sum_{\{a_1,...,a_N\}\in\{a^{(1)},a^{(2)}\}^N}\int_{\R^{2(N+1)}} dp_1 \dots dp_{N+1} d\xi_1 \dots d\xi_{N+1} e^{-\beta H}\delta\left(\sum_{i=1}^N\xi_i+\xi_{N+1}-d\right),
\end{equation}
\end{widetext}
where $\beta=1/k_B T$, $k_B$ is the Boltzmann constant, $T$ the absolute temperature and we have used the Dirac delta in order to include the constraint \begin{equation} d=\sum_{i=1}^N \frac{L}{N}\varepsilon_i+ \alpha L \varepsilon_d
=L (\bar \varepsilon+ \alpha \varepsilon_d)\end{equation}
 fixing the total length $d$. We notice that, due to the structure of the Hamiltonian, the integrals over the momenta are Gaussian and reduce to the constant 
\begin{equation}
A_N=(2 \pi)^{(N+1)/2}\left(\frac{m}{N\beta}\right)^{N/2}\left(\frac{M}{\beta}\right)^{1/2}.
\end{equation} 
Thus, the partition function takes the form
\begin{widetext}
\begin{equation}
Z_N=A_N\sum_{\{a_1,...,a_N\}\in\{a^{(1)},a^{(2)}\}^N}\int_{\R^{N+1}} d\xi_1 \dots d\xi_{N+1} e^{-\beta V}\delta\left(\sum_{i=1}^N\xi_i+\xi_{N+1}-d\right),
\end{equation}

The constraint can be used to perform the integral over $\xi_{N+1}$ so that
\begin{equation}
Z_N=A_N\sum_{\{a_1,...,a_N\}\in\{a^{(1)},a^{(2)}\}^N}\int_{\R^N} d\xi_1 \dots d\xi_{N} e^{-\beta \left( \sum_{j=1}^{N} \Phi(\xi_{i})+\frac{\bar k_d}{2}{(\sum_{i=1}^N\xi_i-d)}^2\right)}.
\end{equation}
\end{widetext}

Finally, we can rewrite the problem in terms of the strains $\varepsilon_i$ using the transformation
\begin{equation}
\xi_i=\frac{L}{N}\varepsilon_i, \quad i=1,\dots N.
\end{equation}
We obtain
\begin{widetext}
\begin{equation}\label{eq:part}
Z_N=C_N\sum_{\{\chi_1,...,\chi_N\}\in\{0,1\}^N}\int_{\R^N} d\varepsilon_1 \dots d\varepsilon_{N}e^{-\beta\frac{lk_p}{2}\left(\sum_i{(\varepsilon_i-\varepsilon_u\chi_i)}^2+\frac{N\gamma}{1-\gamma}{(\delta-\frac{1}{N}\sum_i\varepsilon_i)}^2\right)},
\end{equation}
\end{widetext}
where $C_N=A_N(L/N)^N$, $\varepsilon_u\chi_i=a_i/(L/N)$, $l=L/N$, $\delta=d/L$. Notice that as in the case of zero temperature we have used the main dimensionless parameter $\gamma$ defined in Eq.(\ref{eq:gamma}) of the main paper. We have that
\begin{equation}\label{eq:gammabar}
\gamma=\frac{k_d}{k_d+ \alpha k_p}=\frac{\bar{k}_d}{\bar{k}_d+ \bar{k}_p/N}.
\end{equation}
We notice that  the stiffness of the macromolecule $\bar{k}_{\mbox{exp}}$ measured during an experiment corresponds to the stiffness of $N$ springs in series. Therefore, we have 
\begin{equation}
\bar{k}_{exp}=\bar{k}_p/N=k_p/Nl.
\end{equation}
Thus, we find
\begin{equation}\label{eq:gammabarexp}
\gamma=\frac{\bar{k}_d}{\bar{k}_d+ \bar{k}_p/N}=\frac{\bar{k}_d}{\bar{k}_d+ \bar{k}_{exp}}.
\end{equation}
In the following we will set, without loss of generality, the two minima of the potential wells as $\chi_i=0$ and $\chi_i=1>0, i=1,...,N$. We remark that the choice of  $\bar{k}_{exp}= 4$ pN/nm and $l=24$ nm adopted in the reproduction of Figure {\ref{fig:titin-confront1}} of the main paper is justified by energetic considerations. Indeed, in order to get analytical results, we assumed a parabolic behavior in each well instead {\it e.g.} of a more realistic Worm Like Chain law. Thus, the dissipation associated to each jump is
\begin{equation}
Q=\frac{1}{2}\bar{k}_{exp}\,l^2=1152\;\mbox{pN nm}\simeq 280\, k_B T
\end{equation}
at $300$ K. This is in agreement with the values estimated in \cite{DMPS} based on the experiments performed on PEVK domains \cite{pevk}. 

A simple Gaussian integration of Eq.(\ref{eq:part}) gives
\begin{widetext}
\begin{equation}\label{eq:partitionfunc}
Z_N=C_N{\left(\frac{2\pi }{\beta k_p l}\right)}^{N/2}{\left(1-\gamma\right)}^{1/2}\sum_{\{\chi_1,...,\chi_N\}\in\{0,1\}^N} e^{\frac{\beta k_p l}{2}\left(\sum_i{(\varepsilon_u\chi_i+\frac{\gamma}{1-\gamma}\delta)}^2-\frac{\gamma}{N}{(\sum_i(\varepsilon_u\chi_i+\frac{\gamma}{1-\gamma}\delta))}^2-\sum_i\varepsilon_u^2\chi_i^2-N\frac{\gamma}{1-\gamma}\delta^2\right)}.
\end{equation}
\end{widetext}
The summation over the configurations $\{\chi_1,...,\chi_N\}$ can be rephrased in terms of the fraction $p/N$ of unfolded domains. 
In particular, we have
\begin{equation}
\sum_i\chi_i=\sum_i\chi_i^2=p/N.
\end{equation}
Using this result, we can find the final form of the partition function describing the chain \emph{and} the measuring apparatus:
\begin{equation}
Z_N=K_N{\left(1-\gamma\right)}^{1/2}\sum_{p=0}^N \left( \begin{array}{c} N \\ p \end{array}\right)e^{-\frac{\beta k_p l\gamma N}{2}{(\varepsilon_u\frac{p}{N}-\delta)}^2},
\end{equation}
where 
\begin{equation}
K_N=C_N{\left(\frac{2\pi }{\beta k_p l}\right)}^{N/2},
\end{equation} 
and we have used the binomial coefficient in order to count the number of configurations of the chain with fraction $p/N$ of unfolded domains.

The expectation value of the average strain $\bar{\varepsilon}$ (denoted as $\langle\bar{\varepsilon}\rangle$) can be evaluated as follows. We can use the partition function and the definition of average of a quantity in the canonical ensemble:
\begin{widetext}
\begin{equation}
\langle\bar{\varepsilon}\rangle=\frac{1}{Z_N}C_N\sum_{\{\chi_1,...,\chi_N\}\in\{0,1\}^N}  \int_{\R^{N}} d\varepsilon_1 \dots d\varepsilon_N  \left(\frac{1}{N}\sum_i \varepsilon_i\right) \,e^{-\frac{\beta k_p l\gamma}{2}\left[{\sum_i(\varepsilon_i-\varepsilon_u\chi_i)}^2+\frac{N\gamma}{1-\gamma}{\left(\delta-\frac{1}{N}\sum_i \varepsilon_i\right)}^2\right]}.
\end{equation}
\end{widetext}
It is straightforward to show that
\begin{equation}
\frac{1}{L Z_N}\frac{\partial}{\partial \delta}Z_N=-\beta k_p \frac{\gamma}{1-\gamma} \left(\delta-\langle\bar{\varepsilon}\rangle\right).
\end{equation}
We thus find
\begin{eqnarray}\label{eq:averagestrain}
\langle\bar{\varepsilon}\rangle&=&\delta-\frac{1-\gamma}{k_p \gamma}\left(-\frac{1}{\beta}\frac{1}{L Z_N }\frac{\partial}{\partial \delta}Z_N\right)\nonumber\\ 
&=&\delta-(1-\gamma) (\delta-\varepsilon_u\,\langle \bar \chi \,\rangle )\nonumber\\
&=&\varepsilon_u\,\langle \bar \chi \,\rangle +\gamma\left(\delta-\varepsilon_u\,\langle \bar \chi \,\rangle \right),
\end{eqnarray}
where
\begin{equation}
\langle \bar \chi \,\rangle =\frac{\sum_{p=0}^N \left( \begin{array}{c} N \\ p \end{array}\right)\frac{p}{N} e^{-\frac{\beta k_plN\gamma}{2}{(\varepsilon_u\frac{p}{N}-\delta)}^2}}{\sum_{p=0}^N \left( \begin{array}{c} N \\ p \end{array}\right)e^{-\frac{\beta k_p l N\gamma}{2}{(\varepsilon_u\frac{p}{N}-\delta)}^2}}.
\end{equation}
From these results we obtain Eq.(\ref{eq:finalsigma}) of the paper.

\subsection{Thermodynamical limit}

Let us start considering the sum
\begin{equation}
g(\delta)= \sum_{p=0}^N \left( \begin{array}{c} N \\ p \end{array}\right)e^{-\frac{\beta k_p l\gamma N}{2}{(\varepsilon_u\frac{p}{N}-\delta)}^2}
\end{equation}
which appears in the expression of the partition function (\ref{eq:partitionfunc}).
We can introduce the variable $x=p/N$. In the limit of large $N$, by using the Stirling approximation $n!\simeq (n/e)^n \sqrt{2\pi n}$ for $n\gg 1$, we have
\begin{equation}\label{eq:gfunc}
g(\delta)\simeq\sqrt{\frac{N}{2\pi}}\int_0^1 dx \sqrt{\frac{1}{x(1-x)}}e^{-N \left[S(x)+\tilde{\beta}(\varepsilon_u x-\delta)^2\right]},
\end{equation}
where $\tilde{\beta}= \beta l k_p \gamma /2$ and we have defined the entropy
\begin{equation}
S(x)=x\ln x+(1-x)\ln(1-x).
\end{equation}
In the limit $N\rightarrow+\infty$ we can use the saddle point method and reduce Eq.(\ref{eq:gfunc}) to a Gaussian integral. The explicit calculation gives  
\begin{equation}
g(\delta)\sim \frac{1}{\sqrt{1+2 \tilde{\beta}\varepsilon_u \chi_c (1-\chi_c)}}e^{-N \left[S(\chi_c)+\tilde{\beta}(\varepsilon_u \chi_c-\delta)^2\right]},
\end{equation}
where $\chi_c\in (0,1)$ is the minimum of the function 
\begin{equation}
f(x)=S(x)+\tilde{\beta}(\varepsilon_u x-\delta)^2. 
\end{equation}
An analogous calculation shows that
\begin{eqnarray}
h(\delta)&=&\sum_{p=0}^N \left( \begin{array}{c} N \\ p \end{array}\right)\frac{p}{N}e^{-\frac{\beta k_p l\gamma N}{2}{(\varepsilon_u \frac{p}{N}-\delta)}^2}\nonumber\\
&\sim& \frac{\chi_c}{\sqrt{1+2 \tilde{\beta}\varepsilon_u \chi_c (1-\chi_c)}}e^{-N \left[S(\chi_c)+\tilde{\beta}(\varepsilon_u \chi_c-\delta)^2\right]}.\nonumber \\
\end{eqnarray}
We thus have
\begin{equation}
\langle \bar \chi \,\rangle =\frac{h(\delta)}{g(\delta)}\sim
 \chi_c(\delta),
\end{equation}
and, finally, Eq.(\ref{eq:tlcase1})-(\ref{eq:tlcase2}) of the paper.

\begin{figure}[t]
\begin{center}
\includegraphics[width=0.35\textwidth]{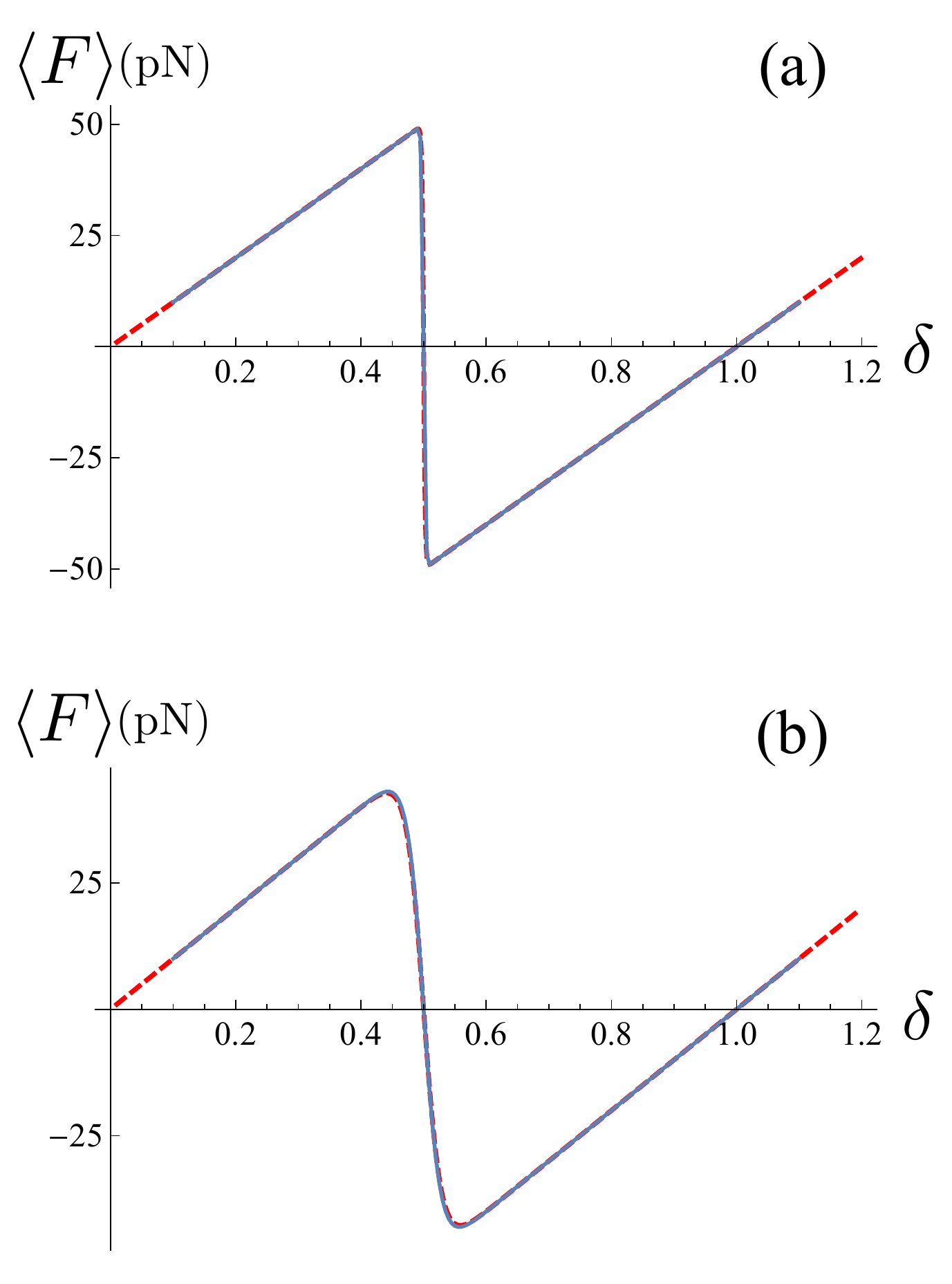}
\caption{Comparison of the force-strain curves obtained using the analytical formula (\ref{eq:sigma}), dashed line, and the numerical integration of the partition function without the approximation beyond the spinodal point described in the text. We have considered two differerent temperatures:(a) $T=300\,\mbox{K}$, (b) $T=3000\,\mbox{K}$. In both cases we have fixed $N=1$ with $l=30 \,\mbox{nm}, \varepsilon_u=1, \bar{k}_d=k_d/(\alpha L)=20\,\mbox{pN/nm}$ and $\bar{k}_p=k_p/l=4\,\mbox{pN/nm}$.}
\label{fig:numeric}
\end{center}
\end{figure}

\begin{figure}[t]
\begin{center}
\includegraphics[width=0.35\textwidth]{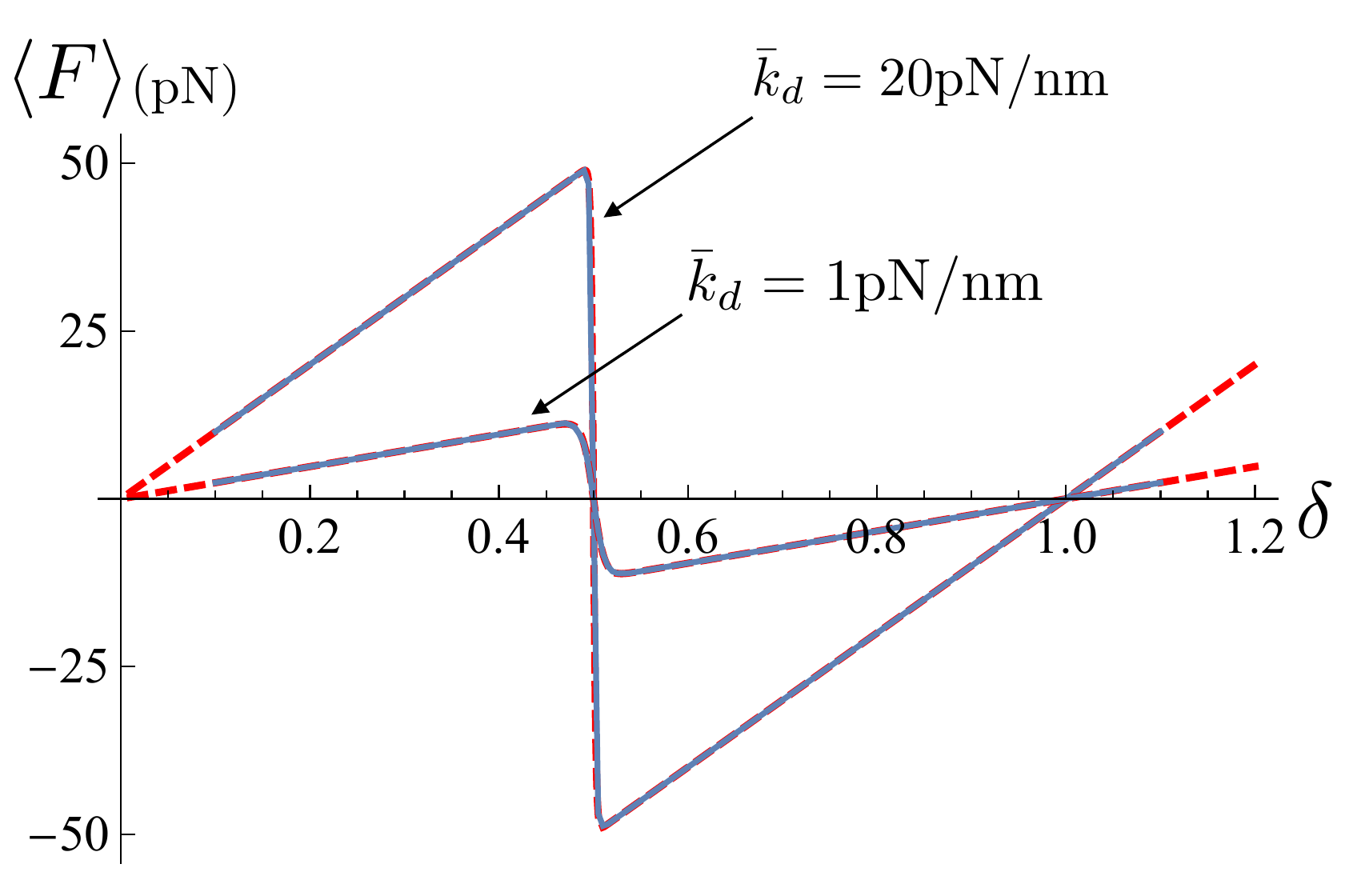}
\caption{Comparison of the force-strain curves obtained using the analytical formula (\ref{eq:sigma}), dashed line, and the numerical integration of the partition function without the approximation beyond the spinodal point described in the text. We have considered two different spring constant of the measurement apparatus. In both cases we have fixed $N=1$ with $T=300\,\mbox{K}, l=30 \,\mbox{nm}, \varepsilon_u=1$ and $\bar{k}_p=k_p/l=4\,\mbox{pN/nm}$.}
\label{fig:numeric2}
\end{center}
\end{figure}

\subsection{Numerical test of the energy approximation}
In this section we include some numerical results in order to confirm the validity of our approximations (extension of the parabolic function beyond the spinodal point) in the chosen range of temperature and values for the parameters of the system. In particular, in Figure \ref{fig:numeric} we compare the force strain curves obtained via Eq.(\ref{eq:sigma}) of the main paper and the numerical integration of the partition function without approximation. Observe that even in the case of very high temperature ($T=3000\mbox{K}$) the approximation is very robust and leads to a result perfectly consistent with the numerical counterpart. Moreover, we test the approximation for different values of the spring constant of the measurement apparatus. The results are shown in Figure \ref{fig:numeric2} supporting the validity of the approximation considered in the paper.


\begin{thebibliography}{00}

\bibitem{ING}
 Ingber, D.E. Mechanobiology and diseases of mechanotransduction.
{\it Ann Med.}  {\bf 35}, 564-77 (2003).

\bibitem{EB}
Essevaz-Rouket, B., Bockelmann, U., Heslot, F, Mechanical separation of the complementary strands of DNA.
{\it 
Proc. Natl. Acad. Sci} {\bf 94}, 11935..40 (1997)

\bibitem{HS} Hummer, G., Szabo, A. Free Energy Surfaces from
Single-Molecule Force
Spectroscopy. {\it Acc. Chem. Res.} {\bf 38}, 504--513
 (2005).

\bibitem{WAS} Woodside, M.T., Anthony, P.C., Behnke-Parks, W.M., Larizadeh, K.,
Herschlag, D., Block, S.M. Direct measurement of the full,
sequence-dependent folding landscape of a nucleic acid.
{\it Science} {\bf 314}, 1001-- 1004 ( 2006).
\bibitem{VT}  Vieregg, J.R., Tinoco, I.J. Modelling RNA folding under mechanical tension. {\it Mol. Phys.} {\bf 104}, 1343-1352  (2006).

\bibitem{MA} Maitra, A.,  Arya G.
Model Accounting for the Effects of Pulling-Device Stiffness in the Analyses of Single-Molecule Force Measurements. {\it Phys. Rev. Lett.} {\bf 104}, 108301 (2010).

\bibitem{MWL}
Manosas,M. ,   Wen, J.D.,  Li, P.T.X., Smith, S.B.,  Bustamante, C.,  Tinoco Jr., I., Ritort F. 
Force Unfolding Kinetics of RNA using Optical Tweezers. II. Modeling Experiments.
{\it Biophys J.} {\bf 92}, 3010--3021 (2007).

\bibitem{LJ} Li, D., Ji, B. Predicted rupture force of a single molecular bond becomes rate independent at ultralow loading rates.
{\it Phys. Rev. Lett.} {\bf 112}, 078302 (2014).

\bibitem{FN} Friddlea, R.W.,  Noyc, A., d, De Yoreoa, J.J.
Interpreting the widespread nonlinear force spectra of intermolecular bonds.
{\it PNAS} {\bf 109}, 13573--13578 (2012)

\bibitem{DHS} Dudko, O,K.,  Hummer, G., Szabo, A.
Theory, analysis, and interpretation of single-molecule
force spectroscopy experiments. {\it PNAS} {\bf 105}, 15755--15760 (2008)

\bibitem{KB} Keten, S., Buehler, M.J. 
Asymptotic strength limit of hydrogen-bond assemblies in proteins at vanishing pulling rates. {\it PRL} {\bf 100}, 198301 (2008)

\bibitem{BMW}
 Bustamante, C., Macosko, J.C.,  Wuite, G.J.L.
Grabbing the cat by the tail: manipulating molecules one by one. {\it Nature Reviews Molecular Cell Biology} {\bf 1}, 130--136 (2000).

\bibitem{KPL} 
Kreuzer, H.J., Payne. S.H., Livadaru L. Stretching a macromolecule in an atomic force microscope: statistical mechanical analysis. {\it  Biophys. J.} {\bf 80}, 2505--2514 (2001).

\bibitem{MGP} Manca, F.,  Giordano,  S.,  Palla, P. L., Zucca, R. , Cleri, F., Colombo L. 
Elasticity of flexible and semiflexible polymers with extensible bonds in the Gibbs and Helmholtz ensembles.
{\it J. Chem. Phys.} {\bf 136}, 154906 (2012). 

\bibitem{PTa} Puglisi, G., Truskinovsky, L. Mechanics of a discrete chain with bi-stable elements.
{\it J. Mech. Phys. Sol.} {\bf 48}, 41--27 (2000).

\bibitem{PTb} Puglisi, G., Truskinovsky, L. Rate independent hysteresis in a bi-stable chain.
{\it J. Mech. Phys. Sol.} {\bf 50}, 165--187 (2002). 

\bibitem{FMG} Florin, E. L.,  Moy, V. T.,  Gaub H. E. Adhesion forces between individual ligand-receptor pairs. {\it Science} {\bf 264}, 415--417 (1994).

\bibitem{Rief} Rief, M., Gautel, M., Oesterhelt, F., Fernandez, J.M.,  Gaub, H.E.  Reversible unfolding of individual titin immunoglobulin domains by AFM. {\it Science} {\bf 276}, 1109 (1997).

\bibitem{CB} 
Cui, Y., Bustamante, C. Pulling a single chromatin fiber reveals the forces that maintain its higher-order structure. {\it Proc. Nat. Acad. Sci.} {\bf 4}, 127--32 (2000).

\bibitem{Ker}  Huang, K. Statistical Mechanics,
John Whiley $\&$ Sons (1987).

\bibitem{martensitic} Falk, F., Konopka, P.
Three-dimensional Landau theory describing the martensitic phase transformation of shape memory alloys. {\it J. Phys.: Condens. Matter} {\bf 2}, 61 (1990).

\bibitem{PT5} Puglisi, G., Truskinovsky, L. Thermodynamics of rate-independent plasticity.
{\it Journal of Mechanics and Physics of solids} {\bf 53}, 655-679 (2005). 

\bibitem{Lev2010} Efendiev, Y.R. , Truskinovsky, L., Thermalization of a driven bi-stable FPU chain, {\it Continuum Mech. Thermodyn. } {\bf 22}, 679 (2010).

\bibitem{Manca} Manca, F. , Giordano, S., Palla, P.L., Cleri, F., Colombo, L. Two-state theory of single-molecule stretching experiments.
{\it Phys. Rev. E} {\bf 87}, 032705 (2013).

\bibitem{NN} Neuman, K.C.,  Nagy A. Single-molecule force spectroscopy: optical
tweezers, magnetic tweezers and atomic
force microscopy. {\it Nature Methods} {\bf 5}, 491--505 (2008).

\bibitem{DMPS} De Tommasi, D.,  Millardi, N.,  Puglisi G.,   Saccomandi G.,
An energetic model for macromolecules unfolding in stretching experiments,
{\it J. R. Soc. Interface} {\bf 10}, 20130651  (2013).

\bibitem{cardiac} Anderson B.R., Bogomolovas J., Labeit S., Granzier H.  Single molecule force spectroscopy on titin implicates immunoglobulin domain stability as a cardiac
disease mechanism. {\it Jour. Biol. Chem.} {\bf 288}, 5303--5315 (2013).

\bibitem{pevk} Linkea W. A., Kulkea M., Lib H., Fujita-Beckerc S., Neagoea C., Mansteinc D.J., Gauteld M., Fernandez J.M. PEVK domain of titin: an entropic spring with actin-binding properties. {\it J. Struct. Biol.} {\bf 137}, 194--205. (2002).

\bibitem{psgl1}
Zhang, Y., Sun, G., Lu, S., Li, N., and Long, M. Low spring constant regulates P-Selectin-PSGL-1 bond rupture. {\it Biophys. Jour.} {\bf 95}, 5439--5448 (2008).

\bibitem{fpt} Truskinovsky, L. Fracture as a phase transition. 
Contemporary research in mechanics and mathematics of materials, Ericksen's Symposium, ed. R. Batra and M. Beatty, CIMNE, Barcelona, pp.322--332 (1996).


\bibitem{goriely} Goriely, A. The Mathematics and Mechanics of Biological Growth, Springer (2017).


\end{thebibliography}
\end{document}